\documentclass{LMCS}

\def\doi{7 (4:07) 2011}
\lmcsheading%
{\doi}
{1--13}
{}
{}
{Mar.~25, 2011}
{Dec.~13, 2011}
{}

\usepackage{hyperref}
\usepackage{url}
\usepackage{enumerate, stmaryrd,mathrsfs}
\usepackage{amsmath,amssymb,latexsym}
\usepackage{xspace} 
\usepackage{pstricks,pst-node}
\usepackage{multido,yhmath}
\usepackage{graphicx}
\usepackage{soul}

\newcommand*{\twodots}{.\,.\,}

\newcommand{\nC}{\newcommand}
\nC{\rnc}{\renewcommand}

\newcommand*{\abgerundet}[1]{\ensuremath{\left\lfloor {#1} \right\rfloor }}

\newcommand*{\set}[1]{\ensuremath{\{ #1 \}}}
\newcommand*{\setc}[2]{\set{#1 \,:\, #2}}
\newcommand*{\dom}[1]{\ensuremath{[#1]}}
\newcommand*{\NN}{\ensuremath{\mathbb{N}}\xspace}   

\newcommand*{\class}[1]{\ensuremath{\textrm{\upshape{#1}}}}
\newcommand*{\FO}{\class{FO}}
\newcommand*{\MonadicNP}{\class{\itshape MonadicNP}}
\newcommand*{\ACzero}{\ensuremath{\class{AC}^0}}
\newcommand*{\ACO}{\ACzero}

\newcommand*{\und}{\ensuremath{\wedge}}

\newcommand*{\oder}{\ensuremath{\vee}}

\newcommand*{\nicht}{\ensuremath{\neg}}
\newcommand*{\impl}{\ensuremath{\rightarrow}}
\newcommand*{\gdw}{\ensuremath{\leftrightarrow}}

\newcommand*{\struc}[1]{\ensuremath{( #1 )}}
\newcommand*{\bigstruc}[1]{\ensuremath{\left( #1 \right)}}
\newcommand*{\A}{\ensuremath{\mathcal{A}}}
\newcommand*{\B}{\ensuremath{\mathcal{B}}}
\newcommand*{\Abig}{\ensuremath{\mathfrak{A}}}
\newcommand*{\Bbig}{\ensuremath{\mathfrak{B}}}
\newcommand*{\abig}{\ensuremath{\mathfrak{a}}}
\newcommand*{\bbig}{\ensuremath{\mathfrak{b}}}

\newcommand*{\praedikat}[1]{\ensuremath{\textsl{#1}}}
\newcommand*{\Bit}{\praedikat{Bit}}
\newcommand*{\Squares}{\praedikat{Squares}}
\newcommand*{\Exp}{\praedikat{Exp}}
\nC{\ord}[1][]{\ensuremath{{<^{#1}}}}
\nC{\ordb}[1][]{\ensuremath{\prec^{#1}}}
\nC{\ordeqb}[1][]{\ensuremath{\preceq^{#1}}}
\nC{\ordc}[1][]{\ensuremath{\prec_{\scriptscriptstyle 0}^{#1}}}
\nC{\ordeqc}[1][]{\ensuremath{\preceq_{\scriptscriptstyle 0}^{#1}}}

\nC{\Max}{\ensuremath{\textit{max}}}
\nC{\Succ}{\ensuremath{\textit{succ}}}
\nC{\Pred}{\ensuremath{\textit{pred}}}
\nC{\phiq}{\ensuremath{\varphi_q}}
\nC{\phir}{\ensuremath{\varphi_r}}
\nC{\phirc}{\ensuremath{\varphi_{rc}}}
\nC{\phibot}{\ensuremath{\varphi_{\text{bot}}}}
\nC{\phisucc}{\ensuremath{\varphi_{\Succ,<}}}
\nC{\phisamecol}{\ensuremath{\varphi_{\text{same-col}}}}
\nC{\phisamerow}{\ensuremath{\varphi_{\text{same-row}}}}
\nC{\phidiag}{\ensuremath{\varphi_{\text{diag}}}}
\nC{\phiQ}{\ensuremath{\varphi_{q,\Bit,r}}}
\nC{\phiR}{\ensuremath{\varphi_{r,\Bit,r}}}
\nC{\phiQRcarry}{\ensuremath{\varphi_{q+r,\text{carry},r}}}
\nC{\phiBitR}{\ensuremath{\varphi_{\Bit,r}}}
\nC{\phiBit}{\ensuremath{\varphi_{\Bit}}}

\begin{document}
\title{A note on the expressive power of linear orders}
\author[N.~Schweikardt]{Nicole Schweikardt\rsuper a}
\address{{\lsuper a}Institut f\"ur Informatik, Goethe-Universit\"at Frankfurt am Main, Germany}
\email{schweika@informatik.uni-frankfurt.de}
\author[T.~Schwentick]{Thomas Schwentick\rsuper b}
\address{{\lsuper b}Lehrstuhl Informatik I, Technische Universit\"at Dortmund, Germany}
\email{thomas.schwentick@tu-dortmund.de}
\begin{abstract}%
 This article shows that there exist two particular linear
 orders such that first-order logic with these two linear orders has the same expressive
 power as first-order logic with the Bit-predicate $\FO(\Bit)$. As a corollary we
 obtain that there also exists a 
 built-in
 permutation such that
 first-order logic with a linear order and this permutation is as
 expressive as $\FO(\Bit)$.
\end{abstract}%
\maketitle
\section{Introduction}\label{section:Introduction}

There are various ways in which arithmetic (i.e., addition and multiplication) on
finite structures can be encoded by other 
numerical
predicates. The
following theorem summarises the results from \cite{BIS90,DDLW98,Lynch82-squares,TroyLee03,Bennett-Phd};
see \cite{Schweikardt_ACMToCL} for a survey. Precise definitions are
given in Section \ref{section:preliminaries}.
\begin{thm}\label{thm:FOBit}
 The following logics have the same expressive power (on the class of all finite structures):
 \begin{quote}
   $\FO(\Bit)$, \quad
   $\FO(<,\Bit)$, \quad
   $\FO(+,\times)$, \quad
   $\FO(+,\Squares)$, \quad
   $\FO(<,\times)$, \\
   $\FO(<,+,\times,\Exp,\Bit,\Squares)$
 \end{quote}
 and each of them can describe exactly those string-languages
 that belong to
 $\class{DLOGTIME}$-uniform $\ACO$.
\end{thm}
From Theorem \ref{thm:FOBit} one might get the impression that
relations with an involved arithmetical structure are necessary to encode arithmetic in a first-order
fashion. Contradicting this intuition, we show in this article that
arithmetic can also be encoded by two particular linear orders. 
More precisely, our main result exposes two
linear orders $\ord,\ordb$ such that $\FO(\ord,\ordb)$ has
the same expressive power as $\FO(\Bit)$. A weaker version of this
result (with three further built-in orders)  had been
announced in \cite{BILST-JCSS,Schweikardt_Diss} (cf., Corollary 5.5(d) 
in \cite{BILST-JCSS} and Theorem 4.5(d) 
in \cite{Schweikardt_Diss}), both referring to an ``unpublished
manuscript on $\MonadicNP$ with built-in grid structures'' by Schweikardt
and Schwentick.
This paper finally presents this
result along with a detailed proof.
As an easy corollary we also obtain a particular built-in permutation
$\pi$ such that $\FO(\ord,\pi)$ has the same expressive power as $\FO(\Bit)$.
\bigskip

\textbf{Organisation.}
The remainder of this paper is structured as follows:  
In Section~\ref{section:preliminaries} our terminology is fixed.
In Section~\ref{section:FO-colgrids} we introduce two linear
orders $\ord,\ordc$ and two unary predicates $C,Q$ and
show that $\FO(\ord,\ordc,C,Q)$ is as expressive as $\FO(\Bit)$.
In Section~\ref{section:cranebeach} we show that
$\FO(<,\ordc)$ is strictly less expressive than $\FO(\Bit)$; the
proof utilises the so-called Crane Beach property that might 
be interesting in its own right.
In Section~\ref{section:linearorders} we show how $\ordc$ and the unary predicates $C$,
$Q$ can be replaced by a single linear order \ordb, and we show how to
represent $\ordb$ by a permutation $\pi$. 
Section~\ref{section:Conclusion} concludes the paper.


\section{Preliminaries}\label{section:preliminaries}%

We write $\NN$ to denote the set $\set{0,1,2,\ldots}$ of all natural
numbers. For each $n\in\NN$ we
write $\dom{n}$ for the set $\set{0,\twodots,n}$ of all natural numbers
of size up to $n$. 
We assume that the reader is familiar with \emph{first-order logic} 
($\FO$, for short), cf., e.g., the textbook \cite{Libkin-FMT}. 

A \emph{$k$-ary numerical predicate} is a relation $P\subseteq\NN^k$.
Particular numerical predicates that were mentioned in the
introduction are
\begin{align*}
  < &:=  \setc{\, (a,b)\in\NN^2\, }{\, a<b\, },\\
  + &:=  \setc{\, (a,b,c)\in\NN^3\, }{\, a+b=c\, },\\
  \times &:=  \setc{\, (a,b,c)\in\NN^3\, }{\, a\cdot b=c\, },\\
  \Squares &:= \setc{\, a\in\NN\, }{\, \text{there exists a }
    b\in\NN \text{ such that } a=b^2\, },\\
  \Exp &:= \setc{\, (a,b,c)\in\NN^3\, }{\, a^b=c\, },\\
  \Bit &:= \setc{\, (a,i)\in\NN^2\, }{\, \text{the $i$-th Bit in the
      binary representation of $a$ is 1, i.e. }
         2 \nmid{\textstyle\abgerundet{\frac{a}{2^i}}}\, }.
\end{align*}
A \emph{$k$-ary built-in predicate} is a sequence
  $(R^n)_{n\in\NN}$ of
  relations, where, for each $n\in\NN$,  $R^n\subseteq \dom{n}^k$.
Clearly, every $k$-ary numerical predicate $P$ naturally induces a $k$-ary built-in predicate
via $P^n := P \cap \dom{n}^k$. 
Note that if $P$ is a strict linear order on $\NN$ (i.e., $P\subseteq\NN^2$ is
transitive, and for all $a,b\in\NN$ we have either $a{=}b$ or $(a,b)\in P$
or $(b,a)\in P$), then $P^n$ is a strict linear order on $\dom{n}$,
for every $n\in\NN$.


\section{Capturing $\FO(\Bit)$ with Two Linear Orders and Two Unary Predicates}\label{section:FO-colgrids}

This section's aim is to present numerical predicates $\ordc$, $C$,
$Q$ such that $\FO(<,\ordc,C,Q)$ captures $\FO(\Bit)$. Here, $C$ and
$Q$ will be unary, and $\ordc$ will be a linear order on $\NN$.

\begin{figure}[ht]
  \centering
\psset{unit=0.5cm}
  \begin{pspicture}(-1.5,-1)(9.5,10)
    
\setlength{\fboxsep}{1pt}

\rput(0,0){0}%
\rput(1,0){1}%
\rput(1,1){2}%
\rput(2,0){3}%
\rput(2,1){4}%
\rput(2,2){5}%
\rput(3,0){6}%
\rput(3,1){7}%
\rput(3,2){8}%
\rput(3,3){9}%
\rput(4,0){10}%
\rput(4,1){11}%
\rput(4,2){12}%
\rput(4,3){13}%
\rput(4,4){14}%
\rput(5,0){15}%
\rput(5,1){16}%
\rput(5,2){17}%
\rput(5,3){18}%
\rput(5,4){19}%
\rput(5,5){20}%
\rput(6,0){21}%
\rput(6,1){22}%
\rput(6,2){23}%
\rput(6,3){24}%
\rput(6,4){25}%
\rput(6,5){26}%
\rput(6,6){27}%
\rput(7,0){28}
\rput(7,1){29}
\psline(-1.5,-0.5)(8.5,-0.5)%
\rput(9.5,-0.5){$\cdots$}%
\psline(-1.5,-1)(7.5,8)%
\rput(8,8.7){$\adots$}%
\rput(-1,0){\fbox{0}}%
 \rput(0,1){\fbox{1}}%
 \rput(1,2){\fbox{2}}%
 \rput(2,3){\fbox{3}}%
 \rput(3,4){\fbox{4}}%
\rput(4,5){\fbox{5}}%
\rput(5,6){\fbox{6}}%
\rput(6,7){\fbox{7}}%
\rput(7,8){\fbox{8}}%
\rput(0,-1){\fbox{0}}%
\rput(1,-1){\fbox{1}}%
\rput(2,-1){\fbox{2}}%
\rput(3,-1){\fbox{3}}%
\rput(4,-1){\fbox{4}}%
\rput(5,-1){\fbox{5}}%
\rput(6,-1){\fbox{6}}%
\rput(7,-1){\fbox{7}}%
\rput(8,-1){\fbox{8}}%
\rput(8.5,0){$\cdots$}%
\rput(8.5,1){$\cdots$}%
\rput(7.5,2){$\cdots$}%
\rput(7.5,3){$\cdots$}%
\rput(7.5,4){$\cdots$}%
\rput(7.5,5){$\cdots$}%
\rput(7.5,6){$\cdots$}%
\rput(7.5,7){$\cdots$}%
%
  \end{pspicture}
  \caption{Illustration of columns and rows for the definition of
    $\ordc$. Row numbers and column numbers are framed.}
  \label{fig:triangle}
\end{figure}
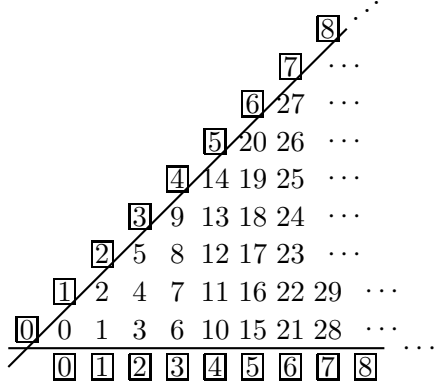

The underlying idea is illustrated in Figure~\ref{fig:triangle}. We
consider the elements of $\NN$ to be distributed into a lower right
triangular matrix with infinitely many columns and rows, where for
every $i\in\NN$, the $i$-th column consists of $i{+}1$ consecutive
numbers, and the $i$-th row contains infinitely many numbers: The
$0$-th column consists of the number 0, the 1-st column consists of
the numbers 1 and 2, the 2-nd column consists of the numbers 3, 4, and
5, and the $i$-th column consists all numbers $z$ with $q_i\leq z \leq
q_i+i$, where $q_i$ denotes the smallest element in this column. I.e.,
$q_0=0$ and $q_{i} = q_{i-1}+i$, for all $i>0$. Thus, $q_i =
\frac{i(i+1)}{2}$, for all $i\in \NN$.

We number the rows from bottom up and the
columns from left to right, starting with 0. 
For each $x\in\NN$, we write $c(x)$ and $r(x)$ to denote the
\emph{column number} and the \emph{row number} of $x$ in
Figure~\ref{fig:triangle}, and we let $q(x)$ denote the bottom-most
element in the same column as $x$. Thus,
\begin{equation}\label{def:functionscrq}
 c(x) =  \max\setc{i\in\NN}{q_i\leq x}, \qquad
q(x) =  q_{c(x)}, \qquad
 r(x) =  x-q(x).
\end{equation}
As an example, $c(13)=4$, $q(13)=10$, and $r(13)=3$. 
Note that, by definition, we have
\begin{equation}\label{eq:nisqnplusrn}
  x=q(x)+r(x) \qquad\text{and}\qquad 0\leq r(x)\leq c(x),  
\end{equation}
for every $x\in\NN$.
Clearly, all numbers $x$ of the same column agree on $q(x)$. We
thus sometimes call $q(x)$ the \textit{$q$-value} of the column of a
number $x$.

Of course, the standard order $<$ on $\NN$ is just the bottom-to-top,
left-to-right, column major order of this matrix. That is, for all
$x,y\in\NN$ we have 
\begin{eqnarray}\label{eq:defOrd}
  x<y & \iff & 
  c(x)<c(y) \quad \text{or}\quad
  \big(\ 
    c(x) = c(y) \ \ \text{and} \ \ r(x) < r(y)
  \ \big).
\end{eqnarray}
\noindent
We define $\ordc$ as the left-to-right, bottom-to-top, row major
order. I.e., for all $x,y\in\NN$ we let
\begin{eqnarray}\label{eq:defOrdc}
  x\ordc y & \iff & 
  r(x)<r(y) \quad \text{or}\quad
  \big(\ 
    r(x) = r(y) \ \ \text{and} \ \ c(x) < c(y)
  \ \big).
\end{eqnarray}
Thus, we have
\[
0\ordc1\ordc3\ordc6\ordc10\ordc\cdots\ordc2\ordc4\ordc7\ordc\cdots\ordc5\ordc8\ordc\cdots\ordc9\ordc\cdots.
\]
\noindent
We
use the relations $C$ and $Q$ to induce binary
  strings on the columns of the matrix. The
  number encoded by the string induced by $C$ on the $i$-th column (with the
bottom-most element of this column representing the least significant
bit) shall\footnote{Why we represent $i+1$, respectively $q_{i+1}$,
  instead of $i$ and $q_i$ will be explained in Footnote~\ref{fn:plusone}.} be $i+1$, and the
  number induced by $Q$ on the $i$-th column shall be $q_{i+1}$.
That is,
\[ C :=  \setc{x\in\NN}{\text{
  bit $r(x)$ of the binary
     representation of $c(x){+}1$ is 1, i.e., }2\nmid\textstyle{\abgerundet{\frac{c(x)+1}{2^{r(x)}}}}},\]
\[ Q := \setc{x\in\NN}{\text{
 bit $r(x)$ of the binary
     representation of $q_{c(x)+1}$ \ is 1, i.e., }2\nmid\textstyle{\abgerundet{\frac{q_{c(x)+1}}{2^{r(x)}}}}}.\]\medskip

\begin{figure}[h!tbp]
 \begin{center}
  \includegraphics[width=6.25cm]{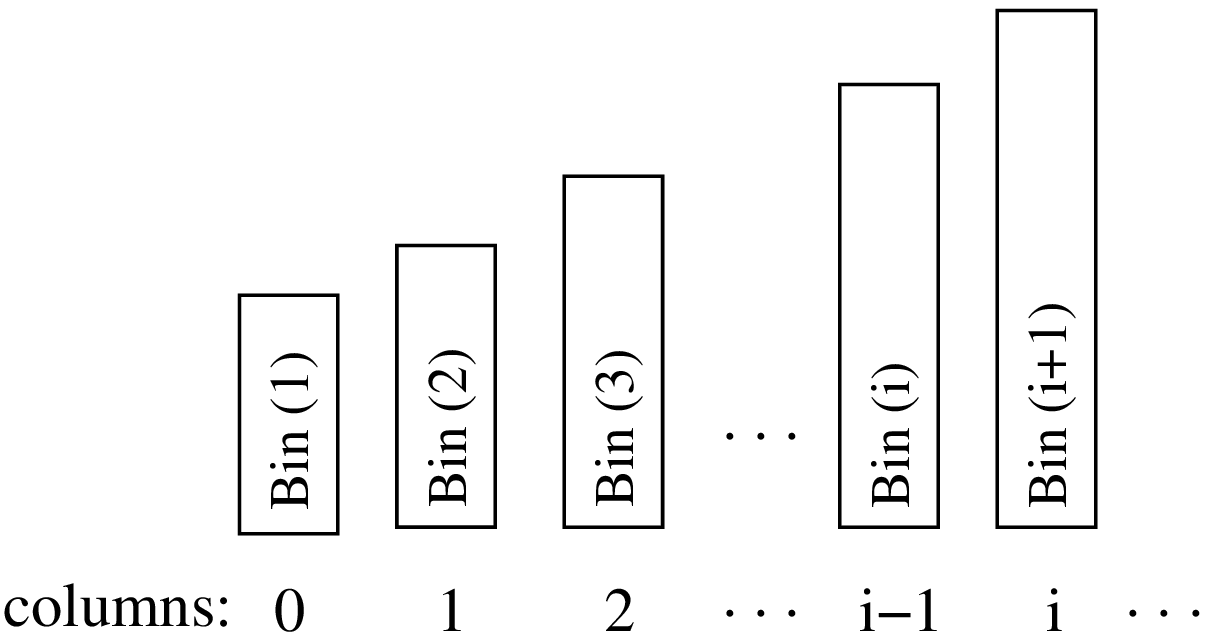}
  \hspace{10mm}
  \includegraphics[width=6.25cm]{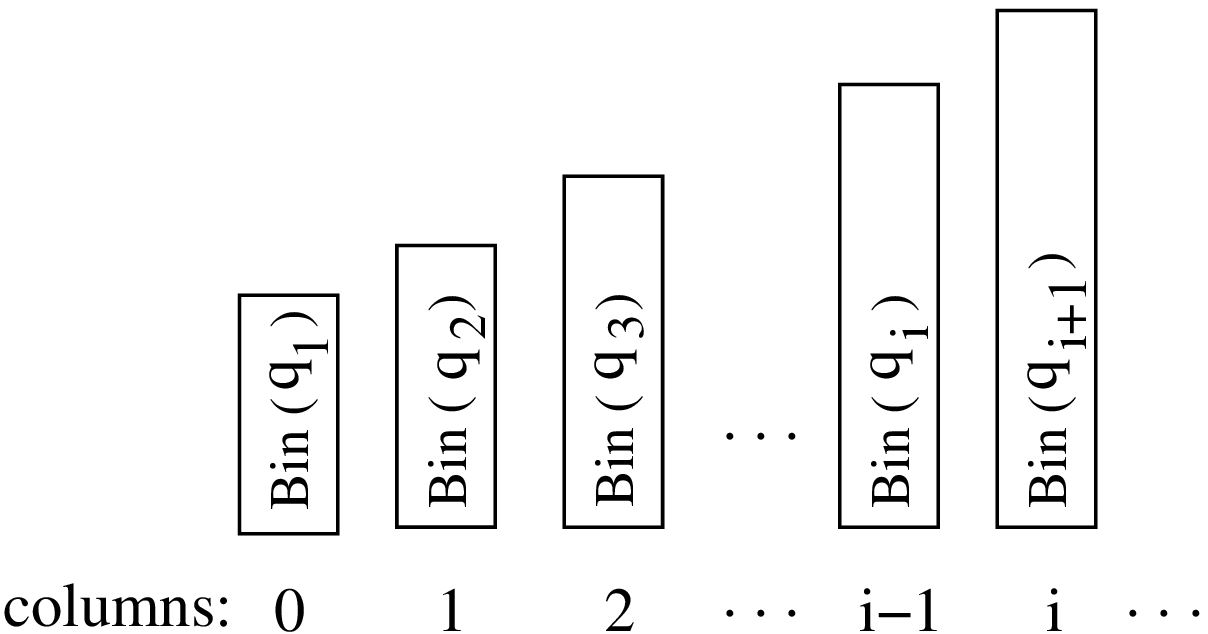}
 \caption{Illustration of the unary predicates $C$ (left) and $Q$
   (right) assigning to each column $i$ the binary representations $\text{Bin}(i{+}1)$ and
   $\text{Bin}(q_{i+1})$ of the numbers $i{+}1$ and $q_{i{+}1}$, respectively. The
   least significant bit of binary representations is in the
   bottom-most row.}
 \label{fig:CQ}
 \end{center}
\end{figure}%
See Figure~\ref{fig:CQ} for an illustration of $C$ and $Q$.
As an example, the restriction of $C$ to column 3 is the set
$\{8\}$ (representing the bit string $100$), and  the restriction of
$Q$ to column 3 is the set $\{7,9\}$ (representing the bit
string $1010$). 

Note that, for every $i$, the $i$-th column contains sufficiently many
elements to encode $q_{i+1}=\frac{(i+1)(i+2)}{2}$, since
the $i$-th column has length $i{+}1$ and can thus encode binary representations of
numbers of size up to $2^{i+1}{-}1 \geq q_{i+1}$.

The remainder of this section is devoted to the proof of the following
theorem.
\begin{thm}\label{thm:mainwithunarypredicates}
$\FO(<,\ordc,C,Q)$ has the same expressive power as $\FO(\Bit)$.
\end{thm}
\proof
That $\FO(\Bit)$ is at least as expressive as $\FO(<,\ordc,C,Q)$
is an immediate consequence of the following
lemma.
\begin{lem}\label{lemma:Bit2ordcetc}
 There are $\FO(\Bit)$-formulas $\varphi_<(x,y)$,
 $\varphi_{\ordc}(x,y)$, $\varphi_C(x)$, $\varphi_Q(x)$ such that, when
 evaluated in $\struc{\dom{n},\Bit^n}$ for some $n\in\NN$, \
 $\varphi_<(x,y)$ expresses that $x<y$, \
 $\varphi_{\ordc}(x,y)$ expresses that $x\ordc y$, \
 $\varphi_C(x)$ expresses that $x\in C$, \ and \
 $\varphi_Q(x)$ expresses that $x\in Q$.
\end{lem}
\proof
The existence of the formula $\varphi_{\ord}(x,y)$
follows from Theorem~\ref{thm:FOBit}.
Using Theorem~\ref{thm:FOBit}, it is straightforward to find $\FO(\Bit)$-formulas
$\varphi_{c}(x,y)$, $\varphi_{r}(x,y)$, and $\varphi_q(x,y)$ which, when interpreted in 
$\struc{\dom{n},\Bit^n}$, express that \ $c(x)=y$, \ $r(x)=y$, and
\ $q(x)=y$, \ respectively.
Using these formulas (and Theorem~\ref{thm:FOBit}), it is an easy
exercise to find formulas
$\varphi_{\ordc}(x,y)$, $\varphi_C(x)$, $\varphi_Q(x)$,
expressing the statement of
equation~\eqref{eq:defOrdc} and the definitions of the predicates $C$
and $Q$.
\qed

To prove the opposite direction, we will construct an
$\FO(<,\ordc,C,Q)$-formula that expresses the $\Bit$-predicate.
The construction of this formula will be established by 
a
sequence of auxiliary formulas and lemmas.

For every $P\in\set{<,\ordc}$ there are $\FO(P)$-formulas
$\varphi_{\Max,P}(x)$ and $\varphi_{\Succ,P}(x,y)$ expressing that $x$
is the maximum element w.r.t.\ the linear order $P$, resp.,
that $y$ is the successor of $x$ w.r.t.\ $P$:
\[
  \varphi_{\Max,P}(x) \ := \ \nicht\exists z\ xPz
  \qquad\text{and}\qquad
  \varphi_{\Succ,P}(x,y) \ := \ \big( xPy \und \nicht \exists z ( xPz \und
  zPy) \big).
\]
For every $c\in\NN$ there is an $\FO(<)$-formula $\varphi_{=c}(x)$
expressing that $x$ is interpreted with the natural number $c$:
\[
  \varphi_{=0}(x) \ := \ \nicht\exists z\ z<x 
  \qquad\text{and}\qquad
  \varphi_{= c+1}(x) \ := \ \exists z\ \big(\varphi_{=c}(z)\und \phisucc(z,x)\big).
\]
To improve readability of formulas, we will henceforth often write
\begin{center}
$x = c$, \quad $x = \Max_P$, \quad $y = \Succ_P(x)$, \quad $y = \Pred_P(x)$ 
\end{center}
instead of $\varphi_{=c}(x)$, \
$\varphi_{\Max,P}(x)$, \ $\varphi_{\Succ,P}(x,y)$, \ $\varphi_{\Succ,P}(y,x)$.
Furthermore, we will write 
\begin{center}
 $x\leq y$ \quad and \quad $x\ordeqc y$ 
\end{center}
as shorthands
for \ $(x<y\,\oder\, x=y)$ \ and \ $(x \ordc y \, \oder \, x=y)$.

\begin{lem}\label{lemma:aux-grid-formulas}
There are formulas 
$\phisamecol(x,y)$, 
$\phisamerow(x,y)$,
$\phiq(x,y)$, and
$\phirc(x,y)$ in $\FO(<,\ordc)$
such that, when evaluated in $\struc{\dom{n},<^n,\ordc[n]}$ for some $n\in\NN$,
\begin{enumerate}[$-$]
\item
 $\phisamecol(x,y)$ expresses that $c(x)=c(y)$, i.e., $x$ is
 in the same column as $y$,
\item
 $\phisamerow(x,y)$ expresses that $r(x)=r(y)$, i.e., $x$ is
 in the same row as $y$, 
\item
 $\phiq(x,y)$ expresses that $q(x)=y$, i.e.,
 $y$ is the bottom-most element in the
 same column as $x$, 
\item
 $\phirc(x,y)$ expresses that $r(x)=c(y)$, i.e.,
 $x$'s row-number is the same as $y$'s column-number.
\end{enumerate}
\end{lem}
\proof
Note that the bottom-most row consists of exactly those elements that
are smaller than 2 w.r.t.\ $\ordc$. Thus we can
choose
\[
  \phibot(x) \ := \ \forall z\ ( z=2 \impl x\ordc z)
\]
to express that $x$ is an element in the bottom row.

Two elements $x$ and $y$ are in different columns iff there exists an
element in the bottom row that lies between $x$ and $y$ w.r.t.\
$<$. Thus, we can choose
\[
  \phisamecol(x,y) \ := \ \nicht\exists z\ \big( \phibot(z)\und
    (x<z\leq y \ \oder \ y<z\leq x)
  \big).
\]
Obviously, 
$q(x)=y$ 
iff $y$ lies in the bottom row and in the same
column as $x$. Thus, we can choose
\[
  \phiq(x,y) \ := \ (\; \phibot(y)\und \phisamecol(x,y)\; ).
\]
For $n\in\NN$ we say that the \emph{last column of $\dom{n}$ is full} 
iff there is an $i\in\NN$ such that $n=q_i+i$. Note that
the last column of $\dom{n}$ is full iff 
$n{=}0$ or the $<$-predecessor of $n$ is also the $\ordc$-predecessor of $n$ and is different from 0. 
This can be expressed by the sentence
\[
\varphi_{\text{last-col-full}} \ := \ \exists z\ \big(
    z=\Max_{<} \ \und \ \big(
         z=0 \ \oder \ \exists y\ 
           (\, y= \Pred_{<}(z)\ \und\  y= \Pred_{\ordc}(z) \ \und\ \nicht\, y{=}0 \,)
      \big)
    \big).
\]
An element $x$ lies on the diagonal (i.e., $r(x)=c(x)$) iff either its $<$-successor lies in the
bottom row, or $x$ is the maximum element w.r.t. $<$ and the last
column is full. Thus, we can choose
\[
  \phidiag(x)\ := \ \big(
    \exists y\ (\,y=\Succ_<(x)\und \phibot(y)\,) \ \oder \ 
    (\, x=\Max_< \, \und \, \varphi_{\text{last-col-full}}\,)
  \big)
\]
to express that $x$ lies on the diagonal.

Two elements $x$ and $y$ lie in different rows iff there exists an
element on the diagonal that lies between $x$ and $y$ w.r.t.\
$\ordc$.
Thus, we can choose
\[
  \phisamerow(x,y) \ := \ \nicht\exists z\ \big( \phidiag(z)\und
    (x\ordc z\ordeqc y \ \oder \ y\ordc z\ordeqc x)
  \big).
\]
Finally, for two elements $x$ and $y$ we have $r(x)=c(y)$ iff the
diagonal element $z$ that is in the same row as $x$, is in the same
column as $y$. Thus we can choose
\[
  \phirc(x,y) \ := \ \exists z\ (\,
   \phidiag(z)\und \phisamerow(x,z)\und\phisamecol(z,y)
   \,).
\]
This completes the proof of Lemma~\ref{lemma:aux-grid-formulas}.
\qed

\begin{lem}\label{lemma:phiQphiR}
There are $\FO(<,\ordc,C,Q)$-formulas $\phiQ(x,u)$ and $\phiR(x,u)$
which, when evaluated in $\struc{\dom{n},<^n,\ordc[n]}$ for some
$n\in\NN$, express that the $r(u)$-th bit of the binary representation
of $q(x)$, respectively, of $r(x)$, is 1.
\end{lem}
\proof
Note that
if $x{=}0$, then $q(x){=}0$, and thus the $r(u)$-th bit of
the binary representation of 
$q(x)$ is 0. If $x>0$, then the binary representation of the number $q(x)$ is given by
relation $Q$ on the elements of the column left to $x$'s column.\footnote{\label{fn:plusone}Here, it is helpful that the $q(x)$ is
  represented in column $c(x)-1$, as this column is guaranteed to be full.} Thus, the $r(u)$-th bit
of $q(x)$ is 1 iff an element $z$ with $r(z)=r(u)$ and $c(z)=c(x)-1$
exists and
belongs to $Q$. Therefore, we can choose
\ $\phiQ(x,u) := $ 
\[
    \exists y \,\exists z\, \big(
          \phisamecol(x,y) \land z {=} \Pred_{\ordc}(y) \land
          \phisamerow(z,y) \land \phisamerow(z,u) \land
           Q(z) 
    \big).
\]
The definition of $\phiQ(x,u)$ is illustrated in Figure~\ref{fig:bit}(a).
\newcommand{\punkt}[3][-45]{\psdot(#2)\uput[#1](#2){#3}}
 \begin{figure}[hbtp]
\psset{unit=0.5cm}
    \centering
    \begin{pspicture}(0,-1)(7,7)
     \rput(4,-1){(a)}
      \psline[linestyle=dashed](0,0)(7,0)
      \psline[linestyle=dashed](0,0)(7,7)
      \psline(7,5)(7,1)
      \psline(2,1)(7,1)
      \punkt{7,5}{$x$}
      \punkt{7,1}{$y$} 
      \punkt{6,1}{$z$}
      \punkt[-135]{2,1}{$u$}
    \end{pspicture}
\hspace{2cm}
    \begin{pspicture}(0,-1)(7,7)
     \rput(4,-1){(b)}
      \psline[linestyle=dashed](0,0)(7,0)
      \psline[linestyle=dashed](0,0)(7,7)
      \psline(7,5)(5,5)
      \psline(5,5)(5,1)
      \psline(2,1)(5,1)
      \punkt{7,5}{$x$}
      \punkt{5,1}{$y$}
      \punkt{4,1}{$z$}
      \punkt[-135]{2,1}{$u$}
    \end{pspicture}
    \caption{Illustration of the meaning of the variables used in (a) $\phiQ(x,u)$ 
      and (b) $\phiR(x,u)$.%
}%
    \label{fig:bit}
  \end{figure}
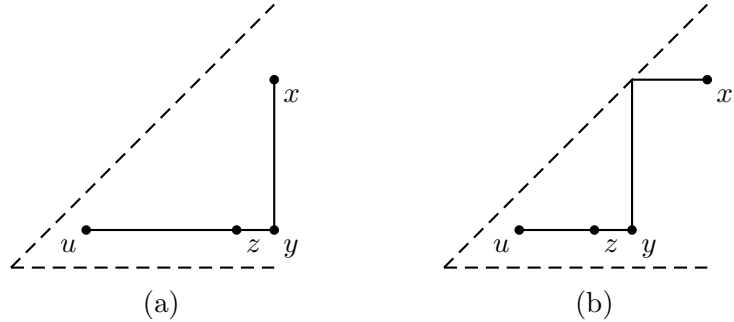

Similarly, if $x{=}0$, then $r(x){=}0$, and thus the $r(u)$-th bit of
the binary representation of 
$r(x)$ is 0. If $x>0$, then the binary representation of the number $r(x)$ is given by
relation $C$ on the elements of the \emph{column} of number $r(x)-1$.
Thus, the $r(u)$-th bit of $r(x)$ is 1 iff an element $z$ with $r(z)=r(u)$ and $c(z)=r(x)-1$
exists and belongs to $C$. Therefore, we can choose
\ $  \phiR(x,u) :=$
\[
   \exists y \,\exists z\; \big(\, \phirc(x,y)\ \land\ 
           z {=} \Pred_{\ordc}(y)  \ \land\, \phisamerow(z,y)\land \phisamerow(z,u) \land
           C(z)\,\big).
\]
The definition of $\phiR(x,u)$ is illustrated in Figure~\ref{fig:bit}(b).
\qed

\begin{lem}\label{lemma:phiBitR}
There is an $\FO(<,\ordc,C,Q)$-formula
$\phiBitR(x,z)$ which, when evaluated in 
$\struc{\dom{n},<^n,\allowbreak \ordc[n]}$ for some
$n\in\NN$, expresses that the $r(z)$-th bit of the binary
representation of $x$ is 1.
\end{lem}
\proof
Recall from equation~\eqref{eq:nisqnplusrn} that $x=q(x)+r(x)$. We
construct the formula $\phiBitR(x,z)$ in such a way that it
expresses that the $r(z)$-th bit in the binary representation of
$q(x)+r(x)$ is 1.

For this, we 
use an auxiliary formula
$\phiQRcarry(x,z)$ which 
expresses that the addition of the binary representations of the numbers
$q(x)$ and $r(x)$ produces a carry-bit to be added at the
$r(z)$-th position.
Note that when adding two binary numbers $a_\ell\cdots a_1 a_0$ and
$b_\ell\cdots b_1 b_0$ (where the least significant bit is at the
rightmost position), a carry-bit has to be added at position $j$ iff
there is a position $i<j$ such that $a_i=b_i=1$ and for all positions
$k$ with $i<k<j$ at least one of the values $a_k,b_k$ is 1. 
Thus, we can choose
\begin{multline*}
\phiQRcarry(x,z)\ := \
\exists u\; \big(\phisamecol(u,z) \ \land \ u< z \ \land \ \phiQ(x,u) \
\land \
\phiR(x,u)\\
\land \ \forall v\; (u<v<z \rightarrow (\phiQ(x,v) \lor \phiR(x,v)))\big).
\end{multline*}

Note
that the $r(z)$-th bit of the binary representation of $q(x)+r(x)$ is 1
if, and only if, 
either no carry-bit has
to be added at position $r(z)$ and the $r(z)$-th bits of $q(x)$ and
$r(x)$ are different, or a carry-bit has to be added at position
$r(z)$ and the $r(z)$-th bits of $q(x)$ and $r(x)$ are the same. Thus,
we can choose 
\[
\begin{array}{ll}
\phiBitR(x,z)\ := &  
 \big(\ \big( \nicht\phiQRcarry(x,z) \ \und \
 (\, \phiQ(x,z)\gdw\nicht\phiR(x,z)\,) \big)\ 
 \oder
 \\
 & \ \ \ \big(\ \ \, \phiQRcarry(x,z) \ \und \ (\,\phiQ(x,z)\gdw
 \ \ \,\phiR(x,z)\,) \big) \ \ \ \ \ \big).
\end{array}
\]
\qed

\begin{lem}\label{lemma:phir}
There is an $\FO(<,\ordc,C,Q)$-formula $\phir(x,y)$ which, when
evaluated in $\struc{\dom{n},<^n,\allowbreak \ordc[n],C^n,Q^n}$ for
some $n\in\NN$, expresses that $r(x)=y$. 
\end{lem}
\proof
Note that $r(x)=y$ iff the following is true:
for every $u$, the $r(u)$-th bit of $r(x)$ is 1 iff the $r(u)$-th bit of $y$ is 1.
We can thus use the formulas $\phiR(x,z)$ and $\phiBitR(y,z)$ from the
Lemmas~\ref{lemma:phiQphiR} and \ref{lemma:phiBitR} to define
\[
 \phir(x,y)\ := \ 
 \forall u\ (\,
   \phiR(x,u) \gdw \phiBitR(y,u)
 \,).
\]
\qed

Now, the $\Bit$-predicate can be expressed by the
$\FO(<,\ordc,C,Q)$-formula stating that there is a number $u$ such that
$r(u)=y$ and the $r(u)$-th bit of $x$ is 1. I.e., we can choose
\[
 \phiBit(x,y)\ := \ \exists u\ (\,
   \phir(u,y) \und \phiBitR(x,u)
 \,).
\]
This finally completes the proof of Theorem~\ref{thm:mainwithunarypredicates}.
\qed


\section{$\FO(<,\ordc)$ Does Not Capture $\FO(\Bit)$}\label{section:cranebeach}

In this section we show that the linear orders $\ord$ and $\ordc$ alone are not sufficient
  to capture $\FO(\Bit)$.
\begin{thm}\label{thm:CraneBeachProperty}
 $\FO(\ord,\ordc)$ is strictly less expressive than $\FO(\Bit)$.
\end{thm}
\proof
Lemma~\ref{lemma:Bit2ordcetc} tells us that $\FO(\ord,\ordc)$
is at most as expressive as $\FO(\Bit)$. 
To show that $\FO(\ord,\ordc)$ does not have the same expressive power as
$\FO(\Bit)$, 
we make use of the so-called \emph{Crane Beach property}
\cite{BILST-JCSS}, which is defined as follows:

\medskip

\begin{enumerate}[$\bullet$]
\item
Let $\ell$ be a list of built-in predicates. The logic $\FO(\ell)$ is
said to have the \emph{Crane Beach property} if the following is true: 
Every string-language $L$ that is definable in $\FO(\ell)$ and that
has a \emph{neutral letter}, is also
definable in $\FO(\ord)$. 
Here, a letter $e$ is called \emph{neutral for $L$}, if 
  for all strings $w_1$, $w_2$ we have \ $w_1w_2\in L \iff w_1ew_2\in L$.
\end{enumerate}

\medskip

\noindent
Clearly, $\FO(<)$ has the Crane Beach property by definition.
From \cite{BILST-JCSS} we know 
that $\FO(\Bit)$ does 
\emph{not} have the \emph{Crane Beach property}.
In the remainder of this proof, we show that $\FO(\ord,\ordc)$ has
the Crane Beach property. This, in particular, will tell us that
$\FO(\ord,\ordc)$ does not have the same expressive power as $\FO(\Bit)$.

The basic idea of the proof that $\FO(\ord,\ordc)$ has the Crane Beach
property is that the order $\ordc$ is useless on structures in which
all columns but the rightmost column contain only neutral letters.
For the proof we follow
the methodology of \cite{BILST-JCSS} and use
\emph{Ehrenfeucht-Fra\"\i{}ss\'{e} games} (EF-game, for short), cf.,
e.g., \cite{Libkin-FMT}. 
Let $L$ be a language that is definable in $\FO(\ord,\ordc)$ and 
that has a neutral letter. 
Let $\Sigma$ be the alphabet of $L$ (i.e., $L\subseteq \Sigma^*$), let
$e\in \Sigma$ 
denote the neutral letter of $L$, and let $k$ be the quantifier
rank  of the $\FO(\ord,\ordc)$-formula that defines $L$. 
Our aim is to show that $L$ is also definable in $\FO(\ord)$. 

Towards a contradiction, let us assume that $L$ is \emph{not} definable in
$\FO(\ord)$. Then, in particular, there are (non-empty) strings $u$ and $v$ such that
$u\in L$, $v\not\in L$, and the duplicator has a winning strategy in
the $2k$-round EF-game
on the structures
\[
  \A \ := \ \bigstruc{\dom{n^u},n^u,\ord[n^u],(Q_\sigma^u)_{\sigma\in\Sigma}}
  \qquad \text{and} \qquad 
  \B \ := \ \bigstruc{\dom{n^v},n^v,\ord[n^v],(Q_\sigma^v)_{\sigma\in\Sigma}}
\]
where, for any string $w$, we let $n^w:=|w|-1$. 
For each letter
$\sigma$ of $\Sigma$ we let 
$Q_\sigma^w := \setc{i\in\dom{n^w}}{w_i=\sigma}$, 
where $w=w_0w_1\cdots w_{n^w}$ with $w_i\in\Sigma$ for all $i\in\dom{n^w}$.
Henceforth, the $2k$-round EF-game on
$\A$ and $\B$ will be called
the \emph{small game}.

Since $L$ has neutral letter $e$, we can assume
without loss of generality that $u$ and $v$ have the same length.
(If not, we can proceed as in \cite{BILST-JCSS}: Append $u$ with $2^{2k}+|v|$
neutral letters $e$, append $v$ 
with $2^{2k}+|u|$ neutral letters $e$, and note that
the duplicator has a winning strategy in the $2k$-round EF-game on the
padded versions of $\A$ and $\B$.)

\begin{figure}[ht]
\begin{minipage}{6cm}
  \centering
\psset{unit=0.5cm}
  \begin{pspicture}(-1.5,-1.5)(9.5,8.5)
    
\setlength{\fboxsep}{1pt}

\rput(0,0){$e$}%
\rput(1,0){$e$}%
\rput(1,1){$e$}%
\rput(2,0){$e$}%
\rput(2,1){$e$}%
\rput(2,2){$e$}%
\rput(3,0){$e$}%
\rput(3,1){$e$}%
\rput(3,2){$e$}%
\rput(3,3){$e$}%
\rput(4,0){}%
\rput(4,1){$\vdots$}%
\rput(4,2){}%
\rput(4,3){$\vdots$}%
\rput(4,4){}%
\rput(5,0){}%
\rput(5,1){$\vdots$}%
\rput(5,2){}%
\rput(5,3){$\vdots$}%
\rput(5,4){}%
\rput(5,5){$\vdots$}%
\rput(6,0){$e$}%
\rput(6,1){$e$}%
\rput(6,2){$e$}%
\rput(6,3){$e$}%
\rput(6,4){}%
\rput(6,5){$\vdots$}%
\rput(6,6){$e$}%
\rput(7.1,0){$u_0$}
\rput(7.1,1){$u_1$}
\rput(7.1,2){$u_2$}
\rput(7.1,3){$u_3$}
\rput(7.1,4){}%
\rput(7.1,5){$\vdots$}%
\rput(7.1,6){}%
\rput(7.1,6.9){$u_{n}$}%

\psline(-1.5,-0.5)(7.6,-0.5)%
\psline(-1.5,-1)(7.6,8.1)%
\psline(7.6,-0.5)(7.6,8.1)%
\rput(-1,0){\fbox{0}}%
 \rput(0,1){\fbox{1}}%
 \rput(1,2){\fbox{2}}%
 \rput(2,3){\fbox{3}}%
 \rput(3,4){\fbox{4}}%
\rput(4.5,5.8){$\adots$}%
\rput(5.8,7.1){\fbox{\large $n$}}%
\rput(7,8){}%
\rput(0,-1){\fbox{0}}%
\rput(1,-1){\fbox{1}}%
\rput(2,-1){\fbox{2}}%
\rput(3,-1){\fbox{3}}%
\rput(4,-1){\fbox{4}}%
\rput(5.5,-1){$\cdots$}%
\rput(7.1,-1){\fbox{\large $n$}}%
  \end{pspicture}
\end{minipage}
\hspace{15mm}
\begin{minipage}{6cm}
  \centering
\psset{unit=0.5cm}
  \begin{pspicture}(-1.5,-1.5)(9.5,8.5)
\setlength{\fboxsep}{1pt}

\rput(0,0){$e$}%
\rput(1,0){$e$}%
\rput(1,1){$e$}%
\rput(2,0){$e$}%
\rput(2,1){$e$}%
\rput(2,2){$e$}%
\rput(3,0){$e$}%
\rput(3,1){$e$}%
\rput(3,2){$e$}%
\rput(3,3){$e$}%
\rput(4,0){}%
\rput(4,1){$\vdots$}%
\rput(4,2){}%
\rput(4,3){$\vdots$}%
\rput(4,4){}%
\rput(5,0){}%
\rput(5,1){$\vdots$}%
\rput(5,2){}%
\rput(5,3){$\vdots$}%
\rput(5,4){}%
\rput(5,5){$\vdots$}%
\rput(6,0){$e$}%
\rput(6,1){$e$}%
\rput(6,2){$e$}%
\rput(6,3){$e$}%
\rput(6,4){}%
\rput(6,5){$\vdots$}%
\rput(6,6){$e$}%
\rput(7.1,0){$v_0$}
\rput(7.1,1){$v_1$}
\rput(7.1,2){$v_2$}
\rput(7.1,3){$v_3$}
\rput(7.1,4){}%
\rput(7.1,5){$\vdots$}%
\rput(7.1,6){}%
\rput(7.1,6.9){$v_{n}$}%

\psline(-1.5,-0.5)(7.6,-0.5)%
\psline(-1.5,-1)(7.6,8.1)%
\psline(7.6,-0.5)(7.6,8.1)%
\rput(-1,0){\fbox{0}}%
 \rput(0,1){\fbox{1}}%
 \rput(1,2){\fbox{2}}%
 \rput(2,3){\fbox{3}}%
 \rput(3,4){\fbox{4}}%
\rput(4.5,5.8){$\adots$}%
\rput(5.8,7.1){\fbox{\large $n$}}%
\rput(7,8){}%
\rput(0,-1){\fbox{0}}%
\rput(1,-1){\fbox{1}}%
\rput(2,-1){\fbox{2}}%
\rput(3,-1){\fbox{3}}%
\rput(4,-1){\fbox{4}}%
\rput(5.5,-1){$\cdots$}%
\rput(7.1,-1){\fbox{\large $n$}}%
  \end{pspicture}
\end{minipage}
\caption{Illustration of the strings $U$ (left) and $V$ (right), embedded in 
    \mbox{$\struc{\dom{N},<^N,\ordc[N]}$}.\newline
    Row and column numbers are framed.}
 \label{fig:cranebeachUV}
\end{figure}
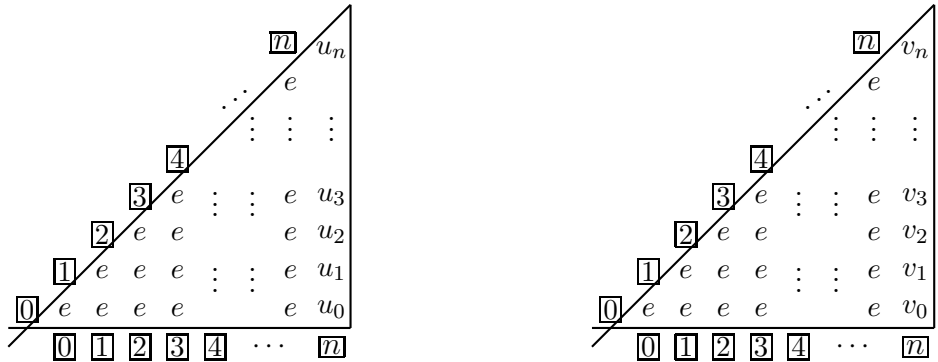

We use $n$ to denote $n^u=n^v$, and we let $u=u_0u_1\cdots u_n$ and
$v=v_0v_1\cdots v_n$ with $u_i,v_i\in\Sigma$.
Now let $N:=q_n +n$, and let $U$ and $V$ be strings of length $N{+}1$ of the form $e^*\,u$
and $e^*\,v$, respectively. 
In particular, we know that $U\in L$ and $V\not\in L$.
Note that $U=U_0U_1\cdots U_N$ is the string which, for all $i$ with $0\leq
i\leq n$ carries letter $u_i$
on position $q_n+i$, and which carries the neutral letter on all other
positions; and analogously
$V$ is obtained from $v$.
An illustration of how $U$ and $V$ are embedded in
$\struc{\dom{N},<^N,\ordc[N]}$ is given in Figure~\ref{fig:cranebeachUV}.

We will now translate the duplicator's winning strategy in the small game into a winning strategy for the $k$-round
EF-game on the structures
\[
  \Abig \ := \ \bigstruc{\dom{N},\ord[N],\ordc[N],(Q_\sigma^U)_{\sigma\in\Sigma}}
  \qquad \text{and} \qquad 
  \Bbig \ := \ \bigstruc{\dom{N},\ord[N],\ordc[N],(Q_\sigma^V)_{\sigma\in\Sigma}}.
\]
Henceforth, the EF-game on $\Abig$ and $\Bbig$ will be called the 
\emph{big game}. Note that $N$, $U$, and $V$ were chosen in
such a way that with respect to the triangular matrix illustrated in
Figure~\ref{fig:triangle} and restricted to the numbers in $\dom{N}$,
the strings $u$ and $v$ are in the 
rightmost column of $\Abig$ and $\Bbig$.

To find a winning strategy for the big game, the duplicator in
parallel plays (according to her given winning strategy) the small
game and translates moves for the small game into moves for the big game.
To be precise,
for every round $i\in\set{1,\ldots,k}$ of the big game, the
duplicator plays two rounds (namely, rounds $2i{-}1$ and $2i$) in the
small game and proceeds as follows: 
If the spoiler chooses an element $\abig_i\in\dom{N}$ in
$\Abig$, the duplicator lets a virtual spoiler choose
$a_{2i-1}:= c(\abig_i)$ and $a_{2i}:= r(\abig_i)$ in the small game
(thus, $\abig_i = q_{a_{2i-1}}+a_{2i}$), considers the
duplicator's answer $b_{2i-1}$ and $b_{2i}$ following her winning
strategy, and chooses
$\bbig_i:= q_{b_{2i-1}}+ b_{2i}$ as her answer in the big
game (thus, $b_{2i-1}=c(\bbig_i)$ and $b_{2i}=r(\bbig_i)$). 
If the spoiler chooses an element $\bbig_i$ in $\Bbig$, the
duplicator's choice of $\abig_i$ in $\Abig$ is determined in the
analogous way.

After the $k$-th round of the big game, we know that the duplicator
has won the small game, since she played according to her winning
strategy. Thus, we have
\begin{enumerate}[(1)]
 \item
   \ $u_{a_i} = v_{b_i}$, \quad for all $i$ with $1\leq i\leq 2k$,
 \item
   \ $a_i < a_j \iff b_i<b_j$, \quad for all $i,j$ with $1\leq i,j\leq 2k$.
\end{enumerate}
Our aim is to show that the duplicator has won the big game, i.e.,
that
\begin{enumerate}[(1')]
 \item
   \ $U_{\abig_i} = V_{\bbig_i}$, \quad for all $i$ with $1\leq i\leq k$,
 \item
   \ $\abig_i < \, \abig_j\ \iff \bbig_i<\,\bbig_j$, \,\quad for all $i,j$ with $1\leq i,j\leq k$,
 \item
   \ $\abig_i \ordc \abig_j \iff \bbig_i \ordc \bbig_j$, \quad for all $i,j$ with $1\leq i,j\leq k$.
\end{enumerate}
Concerning (1'), note that if $a_{2i-1}=n$ then $b_{2i-1}=n$ and 
$\abig_i =q_n+ a_{2i}$, \
$\bbig_i=q_n+b_{2i}$, \
$U_{\abig_i}=u_{a_{2i}}$, \ and
$V_{\bbig_i}=v_{b_{2i}}$. Thus, due to (1) we have
$U_{\abig_i}=V_{\bbig_i}$.
Furthermore, if $a_{2i-1}< n$ then $b_{2i-1}<n$ and
$\abig_i=q_{a_{2i-1}}+a_{2i} < q_n$ and
$\bbig_i=q_{b_{2i-1}}+b_{2i}<q_n$.
Thus, $U_{\abig_i}=V_{\bbig_i}$ is the neutral letter.

To obtain (3'), note that we have
 \[
 \begin{array}{rcll}
   \abig_i\ordc\abig_j 
 & \iff
 & r(\abig_i)<r(\abig_j)\text{ or } \ (r(\abig_i)=r(\abig_j)\text{
      and } c(\abig_i)<c(\abig_j))
 & \text{(by equation~\eqref{eq:defOrdc})}
 \\
 & \iff
 & a_{2i}<a_{2j} \text{ or } \ (a_{2i}=a_{2j}\text{
      and } a_{2i-1}<a_{2j-1})
 & \text{(by def.\ of $a_{2i-1},a_{2i}$)}
 \\
 & \iff
 & b_{2i}<b_{2j} \ \text{ or } \ (b_{2i}=b_{2j}\text{ \
      and } b_{2i-1}<b_{2j-1})
 & \text{(by (2))}
 \\
 & \iff
 & r(\bbig_i)<r(\bbig_j)\text{ or } \ (r(\bbig_i)=r(\bbig_j)\text{
      and } c(\bbig_i)<c(\bbig_j))
 & \text{(by def.\ of $\bbig_i$)}
 \\
 & \iff
 & \bbig_i\ordc\bbig_j
 & \text{(by equation~\eqref{eq:defOrdc})}.
 \end{array}
 \]

Note that (2') can be obtained in the same way, using
equation~\eqref{eq:defOrd}.
\\
In summary, the duplicator has won the big game.
We hence obtain that the structures $\Abig$ and $\Bbig$ satisfy the 
same first-order sentences of quantifier rank $k$. 
However, since $U\in L$ and $V\not\in L$, this
contradicts our assumption that $L$ is definable by an
$\FO(\ord,\ordc)$-sentence of quantifier rank $k$. 
Thus, the proof of Theorem~\ref{thm:CraneBeachProperty} is complete.
\qed


\section{Capturing $\FO(\Bit)$ with Two Linear Orders}\label{section:linearorders}

In this section, we show that in
Theorem~\ref{thm:mainwithunarypredicates} the numerical predicates $\ordc, C, Q$ can be
replaced by one particular linear order. 
The proof will immediately
follow by combining Theorem~\ref{thm:mainwithunarypredicates} with the
following Lemma~\ref{lem:nomonadic}.

If $R,R_1,\ldots,R_k$ are numerical predicates, we say that $R$ is definable in
  $\FO(R_1,\ldots,R_k)$ \emph{in every finite prefix}  if there is an $\FO$-formula that defines $R^n$ on
$(\dom{n},R_1^n,\ldots,R_k^n)$, for every $n\in\NN$. 

\begin{lem}\label{lem:nomonadic}
  For all $k\geq 1$ and 
  all unary relations $U_1,\ldots,U_k$ on $\NN$, there is a linear order
  $\ordb$ on $\NN$, such
  that  $\FO(\ord,\ordb)$ is at least as expressive as
  $\FO(\ord,\ordc,\allowbreak U_1,\ldots,U_k)$ on the class of finite structures. Furthermore, if
  $U_1,\ldots,U_k$ are $\FO(\Bit)$-definable in every finite prefix then $\ordb$
  can be chosen $\FO(\Bit)$-definable in every finite prefix as well.
\end{lem}
\proof
Within this proof, we will use the row numbers, column numbers, and
$q$-numbers defined in equation~\eqref{def:functionscrq}.
Our goal is to encode $\ordc$ and the unary predicates
into a single linear order $\ordb$. 
To this end, the crucial observations are the following:
\begin{enumerate}[(1)]
\item For every number $\ell$, the order $\ordc$ can be recovered in a first-order
  fashion from $<$ and a sub-relation of $\ordc$ that orders only
  every $\ell$-th row (i.e., the rows  $0,\ell,2\ell,\ldots{}$).

\item If $\ell$ is chosen large enough with respect to some  number $m$, the remaining rows allow to
  encode $m$ bits of information per element. 
\end{enumerate}

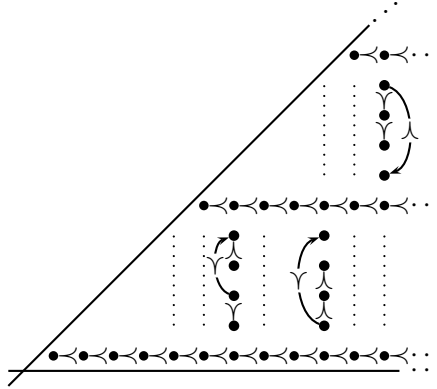
\begin{figure}[ht]
  \centering
\psset{unit=0.4cm}
  \begin{pspicture}(-1,-1)(13,12)
    \psline(-1.5,-0.5)(11.5,-0.5)%
\rput(12.5,-0.5){$\cdots$}%
\psline(-1.5,-1)(10.5,11)%
\rput(11,11.7){$\adots$}%
\multido{\n=0.0+1.0}{12}{\psdot(\n,0)}
\multido{\n=0.5+1.0}{12}{\rput(\n,0){$\ordb$}}
\rput(12.5,0){$\cdots$}
\multido{\n=5.0+1.0}{7}{\psdot(\n,5)}
\multido{\n=5.5+1.0}{7}{\rput(\n,5){$\ordb$}}
\rput(12.5,5){$\cdots$}
\multido{\n=10.0+1.0}{2}{\psdot(\n,10)}
\multido{\n=10.5+1.0}{2}{\rput(\n,10){$\ordb$}}
\rput(12.5,10){$\cdots$}

\cnode*(6,1){2pt}{A}
\cnode*(6,2){2pt}{B}
\cnode*(6,3){2pt}{C}
\cnode*(6,4){2pt}{D}
\rput(6,1.5){\rotateleft{$\ordb$}}
\nccurve[arrows=->,angleA=160,angleB=200]{B}{D}\ncput*[nrot=:U,framesep=0pt]{$\ordb$}
\rput(6,3.5){\rotateright{$\ordb$}}

\cnode*(9,1){2pt}{I}
\cnode*(9,2){2pt}{J}
\cnode*(9,3){2pt}{K}
\cnode*(9,4){2pt}{L}
\rput(9,1.5){\rotateright{$\ordb$}}
\rput(9,2.5){\rotateright{$\ordb$}}
\nccurve[angleA=160,angleB=200,arrows=->]{I}{L}\ncput*[nrot=:U,framesep=0pt]{$\ordb$}

\cnode*(11,6){2pt}{E}
\cnode*(11,7){2pt}{F}
\cnode*(11,8){2pt}{G}
\cnode*(11,9){2pt}{H}
\rput(11,8.5){\rotateleft{$\ordb$}}
\rput(11,7.5){\rotateleft{$\ordb$}}
\nccurve[arrows=->,angleA=-20,angleB=20]{H}{E}\ncput*[nrot=:U,framesep=0pt]{$\ordb$}

\psset{linewidth=0.3mm,linestyle=dotted}
\psline(9,6)(9,9)
\psline(10,6)(10,9)
\psline(4,1)(4,4)
\psline(5,1)(5,4)
\psline(7,1)(7,4)
\psline(10,1)(10,4)
\psline(11,1)(11,4)

  \end{pspicture}
  \caption{Illustration of the definition of $\ordb$ with
    $\ell=5$. For the lower left interval, the corresponding
    permutation $\pi$ is given by $\pi(1)=1$, $\pi(2)=2$, $\pi(3)=4$, $\pi(4)=3$, resulting
    in $u+1\ordb u+2\ordb u+4 \ordb u+3$.}
  \label{fig:nomonadic}
\end{figure}

\noindent
For the given number $k$, we will choose a sufficiently large number $\ell$.
All rows whose number is a multiple of $\ell$ will be called \textit{backbone rows},
and the elements in these rows will be called \textit{backbone elements}.
In $\ordb$, the backbone elements are ordered just
as in $\ordc$, and every backbone element is smaller w.r.t.\ $\ordb$
than every non-backbone element. The number $2$ is
the smallest non-backbone element w.r.t.\ $\ordb$.
Thus, backbone elements can be identified by the $\FO(<,\ordb)$-formula
\[
  \varphi_{\text{backbone}}(x)\ := \ \forall y\ (y=2 \impl x\ordb y).
\]
Figure~\ref{fig:nomonadic} gives an illustration of the overall shape of $\ordb$.

We call a set 
$\{u{+}1,\ldots,u{+}\ell{-}1\}\subseteq\NN$
a \emph{complete interval} if $u$
and $u{+}\ell$ but none of the elements $u{+}1,\ldots,u{+}\ell{-}1$ are backbone elements. In
this case, we call $u$ \emph{complete}. 
We say that $u$ is \emph{complete within $\dom{n}$} if $u$ is complete and $u{+}\ell\in\dom{n}$.
Note that 
there is an $\FO(<,\ordb)$-formula which, when evaluated in
$\struc{\dom{n},<^n,\ordb[n]}$ for some $n\in\NN$, expresses that
$u$ is complete within $\dom{n}$. This formula simply states that
$u$ is a backbone element, $u{+}\ell$ exists, and none of the elements $u{+}1,\ldots,
u{+}\ell{-}1$ is a backbone element.

The elements of complete intervals will be ordered in such a way 
that the order $\ordb$ on every complete interval
$\{u{+}1,\ldots,u{+}\ell{-}1\}$
encodes the unary predicates on the elements
$u$, $u{+}1$, \ldots, $u{+}3\ell{-}1$. 
Note that the encoding is sufficiently redundant to make sure that,
even though there are elements in intervals that are not complete
within $\dom{n}$ (i.e., elements close to the diagonal or close to
$n$), the information
whether $x$ is an element of a set $U_i$ is encoded in \emph{some}
complete interval, for every $x>q_{\ell+1}$.

To describe the order $\ordb$ on each complete interval, we use the following notation.
For every number $x\in\NN$, let $B(x)$ be the bit-string of length
$k$, where the $i$-th bit is 1 if and only if $x\in U_i$. 
For every complete element $u$ we let $\vec{B}(u)$ be the
bit-string of length $3k\ell$ with
\[
   \vec{B}(u) \ := \ \  B(u)\; B(u+1) \cdots B(u+3\ell-1).
\]

We view each bit-string of length $3k\ell$ as the binary
representation of a number from the set
$\{0,1,\ldots,2^{3k\ell}{-}1\}$, and we write $b(u)$
to denote the according number associated with $u$ by the bit-string
$\vec{B}(u)$.
We choose $\ell$ large enough such that 
$(\ell{-}1)!\ge 2^{3k\ell}$. Such an $\ell$
exists, since $n!=2^{\Theta(n\log n)}$ (cf., Stirling's formula) and thus 
$(\ell{-}1)!=2^{\Theta(\ell \log \ell)}$, and hence $(\ell{-}1)!\ge
2^{3k\ell}$ for all sufficiently large $\ell$.
Note that by our choice of $\ell$ we
have $0\leq b(u) \leq (\ell{-}1)!-1$, for every complete element $u$.
 
Let $\pi_0,\ldots,\pi_{(\ell-1)!-1}$ be an enumeration of all permutations
of the set $\{1,\ldots,\ell{-}1\}$. 
Now, the elements of every complete interval $\{u{+}1,\ldots,u{+}\ell{-}1\}$ 
are ordered in $\ordb$ according to $\pi_{b(u)}$ via
\[
  u+\pi_{b(u)}(1)  \ \ \ordb \ \
  u+\pi_{b(u)}(2)  \ \ \ordb \ \
  \cdots \ \ \ordb \ \ 
  u+\pi_{b(u)}(\ell{-}1).
\] 
Note that it is straightforward to construct, for every permutation
$\pi$ of $\set{1,\ldots,\ell-1}$, an $\FO(<,\ordb)$-formula
$\varphi_\pi(u)$ which,
when evaluated in $\struc{\dom{n},<^n,\ordb[n]}$ for some $n\in\NN$,
expresses that $u$ is complete within $\dom{n}$ and the interval
$\{u{+}1,\ldots,u{+}\ell{-}1\}$ is ordered
w.r.t.\ $\ordb$ according to $\pi$. 
 
How 
elements that do not belong to complete intervals, and
how 
elements of different intervals, relate in $\ordb$ does not matter for
our proof. For concreteness, to fully fix $\ordb$, we choose to
let
\[
  x\ordb y \ \iff \  x<y
\]
for all natural numbers $x,y$ for which the relationship has not yet been
defined (neither directly nor transitively).

It remains to verify that
\begin{enumerate}[(a)]
\item the predicates $\ordc$, $U_1,\ldots,U_k$ are 
  $\FO(<,\ordb)$-definable in every finite prefix, and 
\item $\ordb$ is $\FO(\Bit)$-definable in every finite prefix, provided that the unary
  relations $U_1,\ldots,U_k$ are $\FO(\Bit)$-definable in every finite
  prefix. 
\end{enumerate}

Towards (a), we can use the formulas $\varphi_\pi(u)$ to construct, for every
$U_i\in\set{U_1,\ldots,U_k}$, an
$\FO(<,\ordb)$-formula $\varphi_{U_i}(x)$ that, when evaluated in
$(\dom{n},<^n,\ordb[n])$ for some $n\in\NN$, expresses that $U_i(x)$
holds.
Note that either $x< q_{\ell+1}$ or $x=u+j$ where $u$ is an element
complete within $\dom{n}$ and $0\leq j<3\ell$.
In the former case,  the information whether
$U_i(x)$ holds can be ``hard-coded'' into
an $\FO(<,\ordb)$-formula, as $q_{\ell+1}$ is a constant. In the latter
case, the information whether
$U_i(x)$ holds, can be inferred from the particular
permutation $\pi$ for which $\varphi_\pi(u)$ holds.

To express the predicate $\ordc$ by an $\FO(<,\ordb)$-formula, we use that,
for all $x,y\in\NN$, we have
$x\ordc y$  if, and only if,
$x=u+i$ and $y=v+j$ where
$u,v$ are backbone elements and $0\leq i,j <\ell$, such that
the following is true:
\begin{enumerate}[(i)]
 \item 
   $r(u)<r(v)$, \ or
 \item
   $r(u)=r(v)$ \ and \ either $i<j$ or ($i=j$ and $u\ordb v$).
\end{enumerate}
We note that, for backbone elements $u$ and $v$, we have $r(u)<r(v)$
iff 
there is a backbone element $w$ that is the
rightmost element in its row, and $u\ordeqb w \ordb v$. Furthermore, a backbone
element $w$ is rightmost in its row if either it is the maximal backbone element w.r.t.\ to
$\ordb$ or its $\ordb$-successor $w'$ is a backbone element on the
diagonal. The latter can be recognized by the fact that $w'$ and
$w'+1$ are backbone elements.
We can use this to
obtain a formula $\varphi_{\ordc}(x,y)$ expressing that $x\ordc y$.
This concludes (a).

For proving (b) it suffices (due to Theorems~\ref{thm:FOBit} and
\ref{thm:mainwithunarypredicates}) to show that $\ordb$ is
$\FO(\Bit,\ordc,U_1,\allowbreak \ldots,U_k)$-definable in every finite prefix.
 First of all, it is easy to 
identify the backbone rows. Furthermore, it is straightforward (though
tedious) to infer $b(u)$ for a complete element $u$ provided that
$u+3\ell-1\leq n$. 
To infer $b(u)$
for (the at most two) complete elements $u$ with $u+3\ell-1> n$, we use the
fact that, for every $\FO(\Bit)$-formula $\psi(x)$ and every $i\in\NN$
one can construct an $\FO(\Bit)$-sentence $\psi_i$ such that
$(\dom{n},\Bit^n)\models\psi_i$ if and only if
$(\dom{n+i},\Bit^{n+i})\models\psi(n+i)$. 
\qed

From Theorem~\ref{thm:mainwithunarypredicates}, 
Lemma~\ref{lem:nomonadic} and the fact that the predicates $C$ and $Q$
are $\FO(\Bit)$-definable in every finite prefix, we immediately obtain the
main result of this article.
\begin{thm}\label{thm:MainThm-linearorders}
 There is a linear order $\ordb$ on $\NN$ such
  that  $\FO(\ord,\ordb)$ has the same expressive power as
  $\FO(\Bit)$ on the class of all finite structures.
\end{thm}

Using Theorem~\ref{thm:MainThm-linearorders}, one also obtains
the analogous result, where the
linear order $\ordb$ is replaced by a built-in permutation
$\pi=(\pi^n)_{n\in\NN}$, that associates, with every $n\in\NN$, a
permutation $\pi^n$ on the set $\dom{n}$.

\begin{cor}\label{cor:permutation}
There is a built-in permutation $\pi$ such that
$\FO(<,\pi)$ is as expressive as $\FO(\Bit)$.
\end{cor}
\proof
 Let $\ordb$ be the linear order from
 Theorem~\ref{thm:MainThm-linearorders}.
 For every $n\in\NN$ we define $\pi^n$ as follows: For every
 $i\in\dom{n}$ let $\pi^n(i)$
 be the \emph{index} of $i$ w.r.t.\ $\ordb$, i.e., 
 $\pi^n(i) :=  |\setc{j\in\dom{n}}{j\ordb i}|$.
 Then, for all $i,j\in\dom{n}$ the following is true:
 \[
    i\ordb[n] j 
    \ \iff \
    \pi^n(i)\ \ord[n]\ \pi^n(j).
 \]
 Hence, $\ordb$ is definable by the $\FO(<,\pi)$-formula
 $\varphi_{\ordb}(x,y) := \pi(x)\,{<}\,\pi(y)$.
 Therefore, due to Theorem~\ref{thm:MainThm-linearorders},
 $\FO(<,\pi)$ is at least as expressive as $\FO(\Bit)$.

 For the opposite direction, we need to find an $\FO(\Bit)$-formula
 $\varphi_{\textup{index}}(x,y)$ which expresses that $y$ is the index
 of $x$ w.r.t.\ $\ordb$, i.e., $y=\pi^n(x)$. Using our particular
 choice of the linear order $\ordb$ fixed in the proof of Lemma~\ref{lem:nomonadic}, is not difficult
  to construct $\FO(\Bit)$-formulas which
 express that
 \begin{iteMize}{$\bullet$}
 \item $z$ is the total number of backbone elements,
 \item $y$ is the number of backbone elements that are smaller w.r.t.\ $\ordb$
 than  some backbone element $x$, and
\item $y'$ is the number of non-backbone elements that are smaller
  w.r.t.\ $<$ than  some backbone element $x'$. 
 \end{iteMize}
With the help of these formulas the formula
$\varphi_{\textup{index}}(x,y)$ can be constructed. To work out the
 details on the precise definition of this formula is a tedious, but
 easy exercise on $\FO(\Bit)$-definability.
\qed

Let us note that \cite{Schwentick_CSL97} already
exposed a built-in unary function $f$ such that $\FO(<,f)$
has the same expressive power as $\FO(\Bit)$ (see the proof of
Theorem~3 in \cite{Schwentick_CSL97} --- the additional predicate for
multiples of 8 can easily be encoded into $f$). The function $f$
obtained there, however, is not a permutation.


\section{Final Remarks}\label{section:Conclusion}%

We have exposed two linear orders $\ord,\ordb$ and a built-in
permutation $\pi$ such that both,
$\FO(\ord,\ordb)$ and $\FO(\ord,\pi)$ have the same expressive power as $\FO(\Bit)$  
(Theorem~\ref{thm:MainThm-linearorders} and
Corollary~\ref{cor:permutation}).

Of course, it can be debated
  whether linear orders are really ``simpler'' than addition and
  multiplication or the $\Bit$ predicate. Actually, this article
  precisely shows that, with respect to expressive power of
  first-order logic, they are not. However, in an intuitive sense,
  linear orders appear to be simpler, as they are just the transitive
  closure of a linear number of edges, and thus the structure of one
  linear order is more homogenous than, say, the structure of $\Bit$.
The characterisation given in Corollary~\ref{cor:permutation}
even shows that $\FO(\Bit)$ can be captured by using 
$<$ and the linear number of edges provided by the built-in permutation $\pi$.

We note that there is no set $M$ of
\emph{unary} built-in predicates such that $\FO(<,M)$ has at least the
expressive power of $\FO(\Bit)$. This is due to the fact that,
according to \cite{BILST-JCSS}, $\FO(<,M)$ has the Crane Beach
property while $\FO(\Bit)$ does not have this property.

\section*{Acknowledgement} 
We would like to thank Lauri Hella for an inspiring discussion on the
Crane Beach property that led to the proof of Theorem~\ref{thm:CraneBeachProperty}.
Furthermore, we thank the anonymous referees for their valuable comments.


%
\bibliographystyle{plain}
\bibliography{FOGrids} 
\vspace{-100 pt}
\end{document}